\documentclass[a3paper,5p,nofootinbib,superscriptaddress,sort&compress]{elsarticle}

\usepackage[utf8]{inputenc}
\usepackage{epsfig}
\usepackage{booktabs}
\usepackage{footnote}
\usepackage{ctable}
\usepackage{comment}
\usepackage{caption,subcaption}
\usepackage[intlimits]{amsmath}
\usepackage{mathtools}
\usepackage{dcolumn}
\usepackage{bm}
\newcommand {\ra} [1] {\renewcommand{\arraystretch}{#1}}
\newcommand {\beq} {\begin{eqnarray}}
\newcommand {\eeq} {\end{eqnarray}}
\newcommand {\eeqn} [1] {\label{#1} \end{eqnarray}}

\usepackage{multirow}
\usepackage{graphicx}
\usepackage{txfonts}
\usepackage{comment}
\usepackage{amsmath}
\usepackage{breqn}
\usepackage{pifont}
\usepackage{geometry}
\usepackage{fleqn}
\usepackage{textcomp}

\usepackage{hyphenat}

\newcommand\T{\rule{0pt}{2.6ex}}       
\newcommand\B{\rule[-1.2ex]{0pt}{0pt}}

\begin{document}

\title{Unbound neutron $\nu0d_{3/2}$ strength in $^{17}$C and the N=16 shell gap}

\author[USC]{J. Lois-Fuentes}
\author[USC]{B. Fern\'{a}ndez-Dom\'{i}nguez\corref{mycorrespondingauthor}}
\cortext[mycorrespondingauthor]{Corresponding author}
\ead{beatriz.fernandez.dominguez@usc.es}
\author[LPC]{F. Delaunay}
\author[USC,LPC,RAON]{X. Pereira-L\'{o}pez}
\author[LPC]{N.A. Orr}
\author[GANIL]{M. P\l{}oszajczak}
\author[GANIL]{N. Michel}
\author[CNS]{T. Otsuka}
\author[NIH,NAT]{T. Suzuki}
\author[Surrey]{W.N. Catford}
\author[GANIL]{O. Sorlin}
\author[LPC]{N.L. Achouri}
\author[IPNO]{M. Assi\'{e}}
\author[Bham]{S. Bailey}
\author[GANIL]{B. Bastin}
\author[IPNO]{Y. Blumenfeld}
\author[IFINHH]{R. Borcea}
\author[USC]{M. Caama\~{n}o}
\author[GANIL]{L. Caceres}
\author[GANIL]{E. Cl\'{e}ment}
\author[CEA]{A. Corsi}
\author[Bham]{N. Curtis}
\author[LPC]{Q. Deshayes}
\author[GANIL]{F. Farget}
\author[LNS]{M. Fisichella}
\author[GANIL]{G. de France}
\author[IPNO]{S. Franchoo}
\author[Bham]{M. Freer}
\author[LPC]{J. Gibelin}
\author[CEA]{A. Gillibert}
\author[Regina]{G.F. Grinyer}
\author[IPNO]{F. Hammache}
\author[GANIL]{O. Kamalou}
\author[Surrey]{A. Knapton}
\author[Bham]{Tz. Kokalova}
\author[CEA]{V. Lapoux}
\author[US]{J.A. Lay}
\author[IPNO]{B. Le Crom}
\author[LPC]{S. Leblond}
\author[LPC]{F.M. Marqu\'{e}s}
\author[Surrey,LPC]{A. Matta}
\author[IPNO]{P. Morfouace}
\author[US]{A.M. Moro}
\author[GANIL]{J. Pancin}
\author[IPNO]{L. Perrot}
\author[GANIL]{J. Piot}
\author[CEA]{E. Pollacco}
\author[US]{P. Punta}
\author[USC]{D. Ramos}
\author[USC,GANIL]{C. Rodr\'{i}guez-Tajes}
\author[GANIL]{T. Roger}
\author[IFINHH]{F. Rotaru}
\author[LPC]{M. S\'{e}noville}
\author[IPNO]{N. de S\'{e}r\'{e}ville}
\author[Bham]{R. Smith\fnref{myfootnote1}}
\fntext[myfootnote1]{Present address: Department of Engineering and Mathematics, Sheffield Hallam University, Howard Street, Sheffield, S1 1WB, United Kingdom.}
\author[IFINHH]{M. Stanoiu}
\author[IPNO]{I. Stefan}
\author[GANIL]{C. Stodel}
\author[IPNO]{D. Suzuki}
\author[GANIL]{J.C. Thomas}
\author[Surrey]{N. Timofeyuk}
\author[GANIL,CEA]{M. Vandebrouck}
\author[Bham]{J. Walshe}
\author[Bham]{C. Wheldon}

\address[USC]{IGFAE and Dpt. de F\'{i}sica de Part\'{i}culas, Univ. of Santiago de Compostela, 15758, Santiago de Compostela, Spain}
\address[LPC]{Universit\'e de Caen Normandie, ENSICAEN, IN2P3/CNRS, LPC-Caen UMR6534, F-14000 Caen, France}
\address[RAON]{Center for Exotic Nuclear Studies, Institute for Basic Science (IBS), Daejeon 34126, Republic of Korea}
\address[GANIL]{GANIL, CEA/DRF-CNRS/IN2P3, Bd. Henri Becquerel, BP 55027, F-14076 Caen, France}
\address[CNS]{CNS, University of Tokyo, 7-3-1 Hongo, Bunkyo-ku, Tokyo, Japan}
\address[NIH]{Department of Physics, College of Humanities and Sciences, Nihon University, Sakurajosui 3, Setagaya-ku, Tokyo 156-8550, Japan}
\address[NAT]{NAT Research Center, NAT Corporation, 3129-45 Hibara Muramatsu, Tokai, Naka, Ibaraki, 319-1112, Japan}
\address[Surrey]{Department of Physics, University of Surrey, Guildford GU2 5XH, UK}
\address[IPNO]{Universit\'{e} Paris-Saclay, CNRS/IN2P3, IJCLab, F-91405 Orsay, France}
\address[Bham]{School of Physics and Astronomy, University of Birmingham, Birmingham B15 2TT, UK}
\address[IFINHH]{IFIN-HH, P. O. Box MG-6, 76900 Bucharest-Magurele, Romania}
\address[CEA]{D\'{e}partement de Physique Nucl\'{e}aire, IRFU, CEA, Universit\'{e} Paris-Saclay, F-91191 Gif-sur-Yvette, France}
\address[LNS]{INFN, Laboratori Nazionali del Sud, Via S. Sofia 44, Catania, Italy}
\address[Regina]{Department of Physics, University of Regina, Regina, SK S4S 0A2, Canada}
\address[US]{Departamento de FAMN, Facultad de F\'{i}sica, Universidad de Sevilla, Apdo. 1065, 41080 Sevilla, Spain}
%
%

\begin{abstract}

Significant continuum strength has been observed to be populated in $^{17}$C produced in the d($^{16}$C,p) reaction at a beam energy of 17.2~MeV/nucleon. The strength appears at greater than $\sim$2~MeV above the single-neutron decay threshold and has been identified as arising from transfer into the neutron $0d_{3/2}$ orbital.  Guided by shell model predictions the greater majority of the strength is associated with a 3/2$^+$ state at an excitation energy of 4.40$_{-0.14}^{+0.33}$ MeV and a much weaker 3/2$^+$ level at 5.60$_{-0.45}^{+1.35}$ MeV. The corresponding total widths were determined to be 3.45$_{-0.78}^{+1.82}$ and 1.6$_{-1.4}^{+4.6}$ MeV, respectively. From the backward angle proton differential cross sections and the branching ratios for neutron decay to the $^{16}$C(2$_{1}^{+}$) level, the corresponding spectroscopic factors to the ground state were deduced to be 0.47$\pm{10}$ and $<$0.09. Shell-model calculations employing the phenomenological SFO-tls interaction as well as Gamow Shell-Model calculations including continuum effects are in reasonable agreement with experiment, although the predicted strength lies at somewhat lower energy. The size of the N=16 shell gap ($\varepsilon_{ \nu0d_{3/2}}-\varepsilon _{\nu 1s_{1/2}}$) was estimated to be 5.08$_{-0.33}^{+0.43}$~MeV - some 1.3~MeV larger than found in the SFO-tls shell model calculation. 

\end{abstract}

\maketitle

\section{Introduction}

Shell structure is one of the most fundamental characteristics of the nucleus. Indeed, the ordering and relative energy spacing between single-particle orbitals determine the size of shell gaps and the spin-orbit splitting \cite{MayerIII, MayerIV}.  However, the shell gaps and the associated magic numbers are known to evolve as neutrons are added and play an important role in determining, for example, the limits of stability or whether a weakly bound system may exhibit a neutron halo.

Of particular interest in the $p-sd$-shell region is the N=16 shell gap, which is determined by the energy spacing between the $\nu 0d_{3/2}$ and $\nu 1s_{1/2}$ single-particle orbitals, and is directly linked to the location of the neutron dripline for Oxygen, whereby the last bound isotope is doubly-magic $^{24}$O \cite{Hoffman-N16,TshooI}. In this case, the tensor force raises the single-particle $\nu 0d_{3/2}$ orbital towards the neutron $fp$-shell as protons are removed from its spin-orbit partner ($\pi 0d_{5/2}$) creating a gap at N=16 to the detriment of that at N=20 \cite{TOtsukaI,VMU,NSmirnova,SorlinIII}.
Removing two protons from the $0p_{1/2}$ orbital in $^{24}$O, leads to the last bound neutron-rich carbon isotope $^{22}$C which exhibits an s-wave dominated two-neutron halo \cite{Tanaka,Togano,Nagahisa,Kobayashi} and is expected to maintain the large N=16 shell gap \cite{Coraggio}.  In contrast, a recent investigation employing the deformed relativistic Hartree-Bogoliubov model in the continuum (DRHBc model) suggests, based on the agreement with the experimentally deduced $^{22}$C matter radius \cite{Togano}, only a moderately developed halo associated with a quenched N=16 shell gap \cite{Sun}. 

Investigating experimentally the persistence or disappearance of the N=16 shell gap in $^{22}$C is presently beyond the
limits of present facilities.  Mapping the evolution of the single-particle $\nu 0d_{3/2}$ and $\nu 1s_{1/2}$ orbitals
in less exotic C isotopes is, however, possible and can be used to test and improve models and thus their predictions
much further from stability.  It is in this spirit that the present investigation was undertaken with the goal of locating the 
$\nu 0d_{3/2}$ orbital in $^{17}$C.

In the case of $^{15}$C, the ground state exhausts almost all of the $\nu 1s_{1/2}$ strength, whilst the majority of the $\nu 0d_{3/2}$ single-particle strength is found in an unbound state at 4.78~MeV \cite{Darden}.  As a result, the size of the N=16 shell gap, estimated as E$_{x}(3/2^{+})$$-$E$_{x}(1/2^{+})$ = 4.78~MeV, is very similar to that in $^{17}$O -- 4.21 MeV. This persistence of the N=16 gap may be seen as a result of the combined action of the V$^{pn}_{0p_{1/2}0d_{3/2}}$ and V$^{pn}_{0p_{1/2}1s_{1/2}}$ monopole interactions together with continuum effects \cite{Stefan}.
Adding two more neutrons to $^{15}$C could potentially increase the role of neutron-neutron correlations and favour the development of deformation. However, the extent to which the N=16 shell gap persists in $^{17}$C is not known experimentally.  

In terms of the spectroscopy of $^{17}$C, our investigation of the bound states using single-neutron transfer onto $^{16}$C \cite{XPereira} has shown that the neutron $0d_{5/2}$ and $1s_{1/2}$ orbitals are almost degenerate and that the single-particle strengths of the $0d_{5/2}$ and $1s_{1/2}$ orbitals are essentially exhausted \cite{XPereira}. In contrast, there is negligible $0d_{3/2}$ strength found in the bound states\footnote{The 3/2$^+$ ground state exhibits a $^{16}$C(2$^+$)$\otimes$$\nu 0d_{5/2}$ configuration.} and it thus must be located in the continuum.  Similar conclusions have been drawn by other investigations of the ground and bound excited states of $^{17}$C \cite{Ueno,Ogawa,Baumann,SauvanPLB,SauvanPRC,Maddalena,UDatta,MStanoiu-N=14Cchain}.  

Unbound states in $^{17}$C (S$_{n}$= 0.734(18) MeV \cite{AME16}) have been investigated via three-nucleon transfer -- $^{14}$C($^{12}$C,$^{9}$C) \cite{Bohlen}-- inelastic scattering \cite{Satou}, $\beta$-decay \cite{Ueno} and one-neutron removal \cite{Kim}, but in all cases they have been seen to populate mainly negative parity or high spin states. Here, we employ, as in our earlier work \cite{XPereira}, single-neutron transfer onto $^{16}$C using the $^{16}$C(d,p) reaction in inverse kinematics, with the goal of investigating the continuum states of $^{17}$C and locating the $\nu 0d_{3/2}$ strength.

\section{Experiment}

\begin{figure}[h]
\includegraphics[scale=0.45]{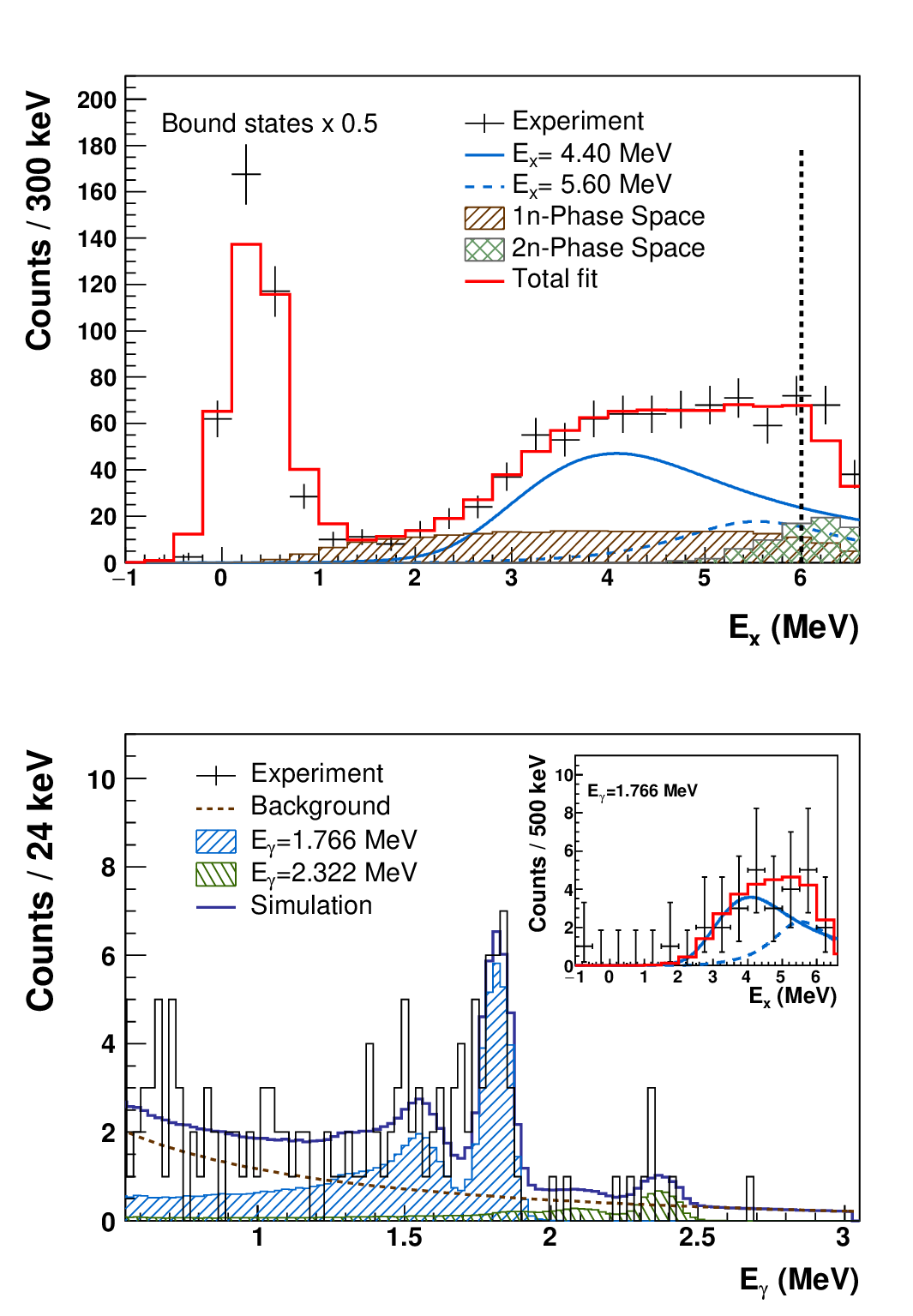}
\caption{\label{data_unbd} \textit{Top panel}: Reconstructed excitation energy of $^{17}$C for $\theta_{lab}=[147^{\circ},167^{\circ}]$ (the vertical error bars are statistical, the horizontal bars represent the binning ).
The overall fit incorporates two resonances at the energies indicated (single-level R-matrix lineshapes convolved with the experimental resolution) and non-resonant contributions from the one and two-neutron breakup channels.
The vertical dotted line shows the limit of the fitting as imposed by the threshold in proton detection (see text). \textit{Bottom panel}:  Addback-reconstructed and Doppler-corrected  $\gamma$-ray energy spectrum for E$_{x}>$ 1~MeV in $^{17}$C. The overall fit is based on the simulated response functions for the  $^{16}$C(2$^{+}_{1}\rightarrow$0$^{+}_{1}$) transition E$_{\gamma}$=1.77 MeV and the (3$^{+}_{1}$, 4$^{+}_{1}$)$\rightarrow$2$^{+}_{1}$ transition E$_{\gamma}$$\approx$2.35 MeV.
A background contribution is also included. \textit{Inset}: $^{17}$C excitation energy for events in coincidence with the $\gamma$-ray peak at 1.77~MeV.}
\end{figure}

A radioactive beam of $^{16}$C at 17.2~MeV/nucleon was produced at the GANIL coupled cyclotron facility 
by fragmentation of a primary beam of $^{18}$O (55~MeV/nucleon) and prepared and purified using the LISE separator \cite{LISE-bis}. The $^{16}$C  intensity at the secondary target position was $\sim$5 $\times$ 10$^{4}$~pps and the purity was essentially 100\%. The secondary target was a CD$_{2}$ foil of thickness 1.37(4) mg/cm$^2$.
The experimental setup around the secondary target was that described in our earlier work \cite{XPereira} - namely the TIARA Si-strip array \cite{TIARA}, 4 Ge clover detectors of the EXOGAM array \cite{EXOGAM} and a Si-Si-CsI CHARISSA telescope \cite{CHARISSA}.  Here, only the data acquired for protons emitted from the $^{16}$C$(d,p)$ reaction at the most backward angles --  $\theta_{lab}$=[147$^{\circ}$,167$^{\circ}$] (corresponding to the annular ``HYBALL'' detector \cite{TIARA}) -- was exploited.  Protons corresponding to excitation energies up to 6~MeV could be measured (Fig. \ref{data_unbd}). At the more forward angles, covered by the detectors of the TIARA  ``Barrel'', much higher thresholds precluded the $^{17}$C continuum from being observed \cite{XPereiraPhD}.

Identification of the atomic number of the beam-like residue was achieved using the CHARISSA telescope placed at zero degrees downstream of the secondary target. The $\gamma$-rays emitted by the beam-like residues were recorded using the EXOGAM detectors  placed at 90$^{\circ}$ and 55 mm from the centre of the secondary target. The absolute photopeak efficiency, including addback and Lorentz boost, was determined to be 6.9(0.2)\% at 1.77 MeV. As will be seen below, the coincident $\gamma$-rays were a key aid in determining the character of the unbound strength. Further details concerning the setup, simulations and data analysis techniques may be found in Refs \cite{XPereira, XPereiraPhD, JLoisPLB, BFD-O21}.

\section{Analysis and Results}

The $^{17}$C excitation energy ($E_x$) was reconstructed from the energy and the angle of the protons detected in coincidence with a Z=6 beam-like residue recorded at zero degrees (Fig. \ref{data_unbd}).  The spectrum shows a clear peak centered at the energy of the second bound excited state (0.335~MeV) and a very broad distribution above E$_{x}\gtrsim$ 2~MeV, that exceeds, as shown in \ref{SM_PS}, the contributions from the non-resonant breakup channels -- $^{16}$C+n+p and  $^{15}$C+2n+p.  

Figure \ref{data_unbd} bottom panel, displays the Doppler-corrected and addback-reconstructed $\gamma$-ray energy spectrum (E$_{\gamma}$ $>$ 500 keV) in coincidence with protons corresponding to $^{17}$C excitation energies in excess of 1 MeV. The $\gamma$-ray energy spectrum was fitted with lineshapes generated using Geant4 \cite{GEANT4}, plus a smooth background \cite{XPereira, XPereiraPhD}. A strong transition is clearly observed at 1.77~MeV and a very weak line at $\sim$2.4 MeV, corresponding, respectively, to the decay of the first $^{16}$C excited state (2$^{+}_{1}$) and from the multiplet of states (2$^{+}_{2}$, 3$^{+}_{1}$, 4$^{+}_{1}$) at around 4.1 MeV to the 2$^+_{1}$ state \cite{ENSDF}. Energetically both resonances in $^{17}$C can decay to the first 2$^{+}$ state in $^{16}$C and, despite the limited statistics, it is clear that both states do so -- inset of Figure\ref{data_unbd} (bottom panel). 
Taking into account the $\gamma$-ray and proton detection efficiencies, the first resonance was found to decay to  $^{16}$C(2$^{+}_{1}$ ) with a branching ratio, BR$_{1}$(2$^{+}_{1}$), of 0.66(15), and the corresponding branching ratio, BR$_{2}$(2$^{+}_{1}$), of the second resonance was found to be 1, with a $1\sigma$ uncertainty lower limit of 0.72.  In each case, any contribution from feeding from neutron-decay to the higher-lying multiplet is smaller than the uncertainties.

Guided by shell model calculations, such as those presented below, which suggest that a number of levels should lie in the region of $\sim$2--6~MeV, the excitation energy spectrum was described in terms of two broad resonances (found to be located at 4.4 and 5.6~MeV excitation energy) and contributions from the non-resonant single and two-neutron breakup channels\footnote{These will also take account for any contributions from very broad higher-lying states.}. Importantly,  the description of the continuum in terms of a single resonance (\ref{SM_1R}) does not change the essential conclusions of the present work.  Indeed, in the two-resonance description, the great majority of the strength is carried by the lower lying level. 

The resonances were described in terms of single-level R-matrix lineshapes, 

\begin{equation}
       \sigma_{r}(E_x,E_{0}) \propto \frac{\Gamma^{tot}\left(E_{x}\right)}{ \left( E_{x}-E_{0}\right)^{2}+\left(\Gamma^{tot}\left(E_{x}\right)/2\right)^{2}}
\end{equation}

where $E_0$ corresponds to the excitation energy of the resonance.  The total width of the resonance is a combination of the widths of the two decay channels 0$^{+}_{1}$ and 2$^{+}_{1}$ (see discussion below),

\begin{equation}
\Gamma^{tot}(E_x,E_0) = \Gamma_{\ell}(0^{+}_{1}) + \Gamma_{\ell}(2^{+}_{1})
\end{equation}

\begin{equation}
\Gamma_{\ell}(E_x,E_0) = \Gamma_{r} \frac{P_{\ell} (E_{x})}{ P_{\ell} (E_{0})}
\end{equation}

where  $P_{\ell}$ is the penetrability for a neutron with orbital angular momentum $\ell$ \cite{LaneThomas}.

The resonance lineshapes were convoluted with the resolution in excitation energy (FWHM$=$705~keV) which was determined using a Monte Carlo simulation \cite{XPereiraPhD} based on Geant4 \cite{GEANT4}, that included the beam characteristics (spot size and energy spread), the energy and angular straggling in the target, the Si-detector resolutions and kinematical effects.  
The amplitudes, energies and widths of the resonances were free parameters and only the experimental resolution was fixed in the fitting procedure. 
The non-resonant contributions arising from breakup were estimated by sampling, to take into account the experimental acceptances, the simulated n-body phase space distributions.  In the case of the 1n breakup channel, the amplitude of its contribution is strongly constrained by the yield in the excitation energy spectrum between 1 and 2~MeV.
The peak associated with the closely spaced bound states was described using a Gaussian distribution, with a width governed by the known experimental resolution, and an amplitude which was left to vary in the fit.

As alluded to above, no strongly populated 1/2$^+$ or 5/2$^+$ unbound states are expected as the bound states with these spin-parties carry most of the expected strength.  In the case of a 3/2$^-$ state with all but the smallest spectroscopic factor (and hence population), the associated width would be far too broad to allow it to be distinguished from the non-resonant continuum.  As such, the unbound strength observed here, in transfer of a neutron onto the $^{16}$C ground state is most probably associated with J$^\pi$=3/2$^+$ and/or 7/2$^-$ levels.  Shell model calculations suggest, for a range of interactions \cite{Kim}, that the former are more likely to dominate the excitation energies populated here with the latter lying higher in energy.

\begin{figure}[h]
\includegraphics[scale=0.35]{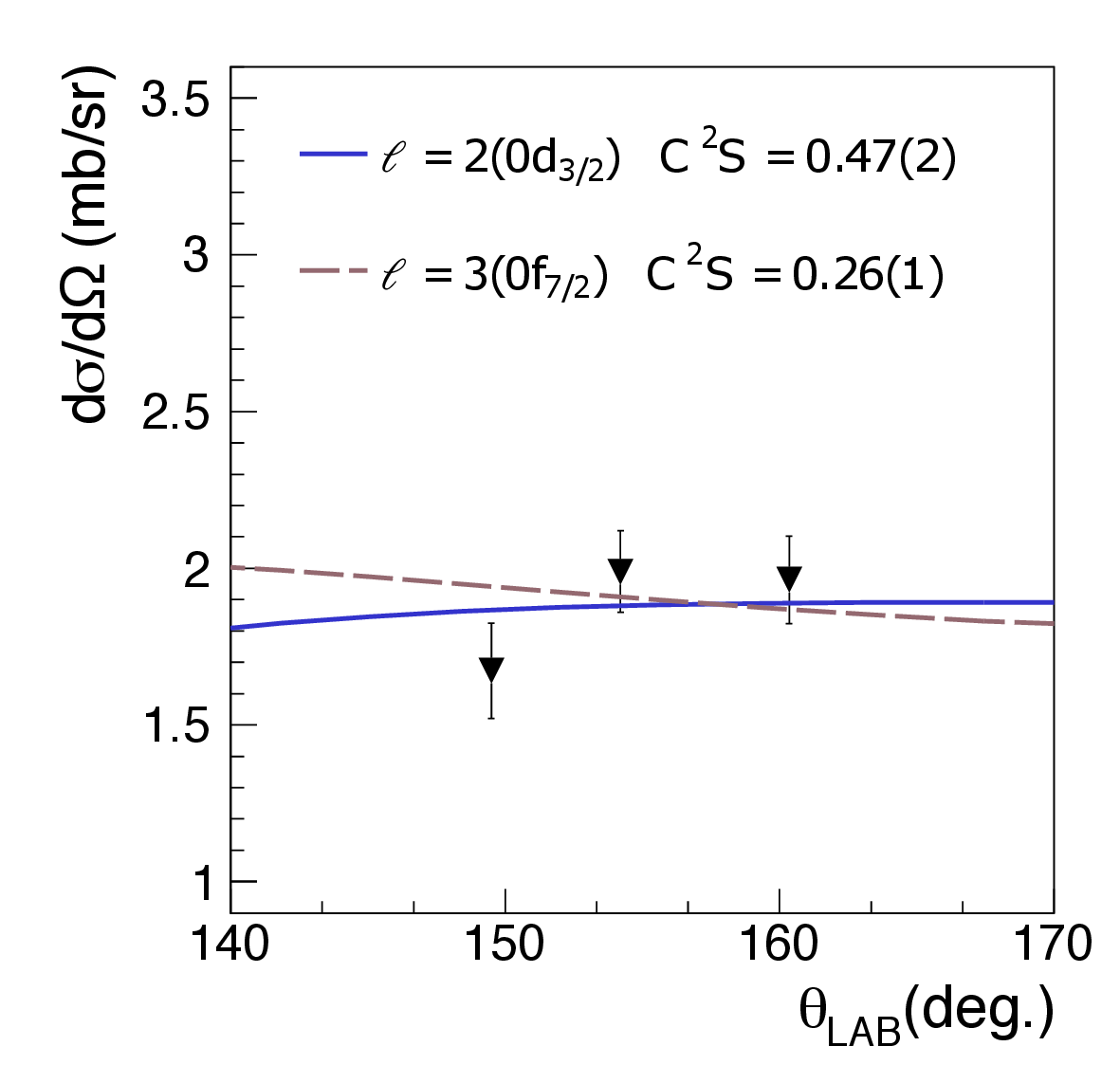}
\caption{\label{ang_dist} Proton backward angle differential cross sections for the lower-lying resonance (data points shown with statistical error bars) compared to ADWA calculations for transfer to the $\nu0d_{3/2}$ (solid blue line) and $\nu0f_{7/2}$ (dashed brown line) orbitals.}
\end{figure}

Owing to the restricted range of angles covered by the present data and the very broad character of the resonances, it was not possible to deduce the transferred angular momentum from the proton differential angular distributions. However, the backward angle differential cross section (Fig.~\ref{ang_dist}) can be used to deduce the spectroscopic factor for the lower-lying resonance (for the much weaker higher lying resonance the statistics was very limited). Specifically, the experimental distribution is compared to Adiabatic Distorted Wave Approximations (ADWA) calculations computed with DWUCK4 \cite{dwuck} using the Koning-Delaroche nucleon-nucleus optical potential parameterisation \cite{KoningDelaroche} in the exit channel, as well as in the entrance channel to compute the $d+^{16}$C adiabatic potential.
Unbound $^{16}$C+n form factors were computed in a Wood-Saxon potential with standard geometry (radius parameter $r_0=1.25$~fm and diffuseness $a=0.65$~fm) and with the depth adjusted to give a resonance at the measured energy.  Transfer to these unbound form factors was considered using the Vincent and Fortune procedure \cite{vf}. The calculations employed the zero-range approximation with a standard finite-range correction and a standard non-locality correction in the exit channel.
Spectroscopic factors were obtained by normalising the computed differential angular distribution cross sections to that measured (Fig.~\ref{ang_dist}).  As
noted above, no 3/2$^-$ ($\ell$=1) resonance can be formed at the energies in question here, and comparison is thus made with $\ell$=2 ($0d_{3/2}$) and 3 ($0f_{7/2}$) neutron transfer onto the $^{16}$C$(0^+_1)$ ground state and corresponding spectroscopic factors of
 $C^{2}S_{1}$(0$^{+}$)=0.47(2) and 0.26(1) were deduced, where the uncertainty is only statistical\footnote{As discussed in earlier studies, an uncertainty of some $\pm$20\% associated with the reaction modelling is expected - see Ref.~\cite{XPereira} and refs therein.}.

\begin{table*}
\centering
\ra{1.4}
\caption{Results for the fitting of the lower-lying resonance (see text): excitation and resonances energies, $E_{x}$ and $E_{r}$ (MeV), spin and parity (J$^{\pi}$, valence neutron quantum numbers $n\ell j$, total and partial widths $\Gamma$ (MeV) and spectroscopic factors $C^2S$.  Single-particle widths for $\ell=2$ and $3$ were calculated using a Woods-Saxon potential with standard geometry (radius parameter $r_0=1.25$ fm and diffuseness $a=0.65$ fm), with the depth adjusted to the resonance energy.}
\vspace{0.15cm}

\tabcolsep=0.11cm

\begin{tabular}{cccccccccccc}
\hline
\hline
$E_{x}$ & ${E_{r}^{a)}}$ & $J^{\pi}$ &$n \ell j ~ \otimes$ 0$^{+}$ &  $n \ell j ~ \otimes$ 2$^{+}$ & $ \Gamma^{tot}_{1}$  &   $ \Gamma_{1}$(0$^{+}$)  &  $ \Gamma_{sp}$(0$^{+}$)  & $C^{2}S_{1}(0^{+})$   &   $ \Gamma_{1}$(2$^{+}$)  &  $ \Gamma_{sp}$(2$^{+}$)  & $C^{2}S_{1}(2^{+})$  \\
\hline
4.40$^{+0.33}_{-0.14}$  & 3.66$^{+0.33}_{-0.14}$ & 3/2 $^{+}$ & $0d_{3/2}$ & $1s_{1/2}$  & 3.45$_{-0.78}^{+1.82}$  & 1.17$_{-0.58}^{+0.81}$ &  2.58  &  0.45$_{-0.22}^{+0.32}$   & & $^{b)}$  &  \T \\
4.32$^{+0.19}_{-0.21}$  & 3.58$^{+0.19}_{-0.21}$ & 3/2 $^{+}$ & $0d_{3/2}$ & $0d_{3/2}$  & 3.05$_{-0.64}^{+0.91}$  & 1.04$_{-0.51}^{+0.55}$ &  2.44  &  0.43$_{-0.58}^{+0.81}$  & 2.01$_{-0.62}^{+0.75}$ & 0.518 & 3.9$_{-1.2}^{+1.5}$ \\

4.41$^{+0.72}_{-0.23}$ &  3.67$^{+0.72}_{-0.23}$ &  7/2 $^{-}$ & $0f_{7/2}$ & $1p_{3/2}$  & 1.84$_{-0.21}^{+2.90}$  & 0.63$_{-0.29}^{+1.02}$ &  0.405  &  1.56$_{-0.71}^{+2.51}$  &  & $^{b)}$ & \\
4.41$^{+1.04}_{-0.07}$  & 3.67$^{+1.04}_{-0.07}$ & 7/2 $^{-}$ &$0f_{7/2}$ & $0f_{7/2}$  & 1.89$_{-0.21}^{+5.99}$  & 0.64$_{-0.29}^{+2.06}$ &  0.405  &  1.58$_{-0.71}^{+5.08}$   &  1.25$_{-0.31}^{+3.96}$ & 0.052 & 24 $_{-6}^{+76}$ \B \\
\hline
\hline
\label{sf_exp}
\end{tabular}
 
\end{table*}%
\begin{table*}
\vspace{-0.7 cm}
\begin{tabular}{@{}p{15.9 cm}@{}}
 \hspace{1.3 cm}$^{a)}$ \footnotesize{For the decay to the $^{16}$C ground state}\\
 \hspace{1.3 cm}$^{b)}$ \footnotesize{No resonance possible at corresponding decay energies.}\\
\end{tabular}
\end{table*}

We now turn to the structural information which may be obtained from the widths for the
neutron decay to the $^{16}$C(0$^+_1$) ground and $2^+_1$ states and the measured branching ratios.
For the higher-lying resonance, the very large branching ratio for neutron decay to the $^{16}$C$(2^+_1)$ state (BR$_2(2^+_1) \simeq 1$), indicates that its structure is dominated by core excitations with a structure compatible, given that only a J$^{\pi}$=3/2$^{+}$ or 7/2$^{-}$ assignment is possible (see above), with a $d-$wave neutron -- very probably $d_{3/2}$ -- coupled to the $^{16}$C$(2^+_1)$, in line with the shell-model calculations presented below for the higher-lying strength.  A J$^{\pi}=7/2^{-}$ assignment implies coupling to either a $p_{3/2}$ neutron, which will form an extremely broad structure or an $f_{7/2}$ neutron which will generate a resonance far too narrow compared to the observed width ($\sim$2~MeV) which is dominated by the decay to the $^{16}$C$(2^+_1)$ state.

In the case of the lower-lying resonance the neutron decay strength is shared between that to the $^{16}$C$(2^+_1)$ $-$ BR$_{1}(2^{+}_{1}$) = 0.66(15) $-$ and the feeding of the $^{16}$C$(0^+_1)$ ground state - BR$_{1}$(0$^{+}_{1}$) = 0.34(15).  
Fits of the excitation energy spectrum were carried out using the R-matrix lineshapes together with
the contributions from the non-resonant continuum, as described earlier.
In the case of the lower-lying resonance these were performed for the configurations possible for J$^\pi$=3/2$^+$ and 7/2$^-$
assignments as listed in Table~\ref{sf_exp}, whilst the higher-lying resonance was included with the $\nu0d_{3/2}$$\otimes$$^{16}$C$(2^+_1)$ configuration\footnote{Within the relatively large uncertainties associated with the fitting, the results for the lower-lying resonance were insensitive to the characteristics of the higher-lying resonance.}.
It should be noted that whilst the energy of the lower-lying resonance is not sensitive, within uncertainties to the spin-parity and associated configurations, the total and partial decay widths depend clearly on the spin-parity.
 
Given the branching ratios for the neutron decay channels (BR$_{1}(0^{+}_{1}$) and BR$_{1}(2^{+}_{1}$)) the associated spectroscopic factors, $C^{2}S$, may be deduced based on the corresponding single-particle widths, $\Gamma_{sp}$.  
Most significantly, as may be seen in Table~\ref{sf_exp}, only the cases corresponding to a J$^\pi$=3/2$^+$ assignment exhibit $C^{2}S_{1}(0^{+})$ which are in accord with the value deduced from the proton backward angle differential cross sections (Fig.~\ref{ang_dist}). 
In terms of the component of the lower-lying resonance associated with the $^{16}$C$(2^+_1)$, the spectroscopic factor,
$C^{2}S_{1}(2^{+}_{1} )$, associated with a $0d_{3/2}$ neutron is unphysical large.  Whilst we cannot compute a single-particle width for the $1s_{1/2}$ neutron, it may be noted that the shell-model calculations (see below and Table~\ref{tab_sf_17c}) favour such a configuration for the predicted lower-lying strength.

With both resonances identified as 3/2$^+$ with the configurations and branching ratios described, the best-fit of the excitation energy spectrum (Fig.~\ref{data_unbd}) yielded energies and total widths of E$_{x,1}$ = 4.40$_{-0.14}^{+0.33}$, $\Gamma_{1}^{tot}$=3.45$_{-0.78}^{+1.82}$ and E$_{x,2}$ = 5.60$_{-0.45}^{+1.35}$, $\Gamma_{2}^{tot}$=1.6$_{-1.4}^{+4.6}$ MeV. As such, for the higher-lying resonance, given the upper limit for the decay to the $^{16}$C$(0^+_1)$ ground state (BR$_{2}(0^{+}_{1}$) $<$ 0.28) and the corresponding single-particle width for a d$_{3/2}$ neutron ($\Gamma_{sp}$=5.16~MeV), an upper limit for the spectroscopic factor could be deduced -- $C^{2}S_{2}(0^{+})<0.09$.

Finally, as indicated earlier, the unbound strength is dominated by that of the lower-lying resonance.   As such, although it is at some variance to the model predictions (see following) the excitation energy spectrum can also be well described, as shown in \ref{SM_1R}, by a fit incorporating only a single much broader 3/2$^+$ resonance with E$_{x,1}$ = 5.17$_{-0.43}^{+1.53}$  and $\Gamma_{1}^{tot}$=6.2$_{-1.9}^{+8.0}$ MeV. Importantly, this does not change, in any significant manner the 3/2$^+$ spectroscopic strength nor the energy deduced for the effective single-particle energy for the $\nu 0d_{3/2}$ orbital and the N=16 shell gap.

\section{Discussion}

\begin{figure}
\includegraphics[scale=0.5]{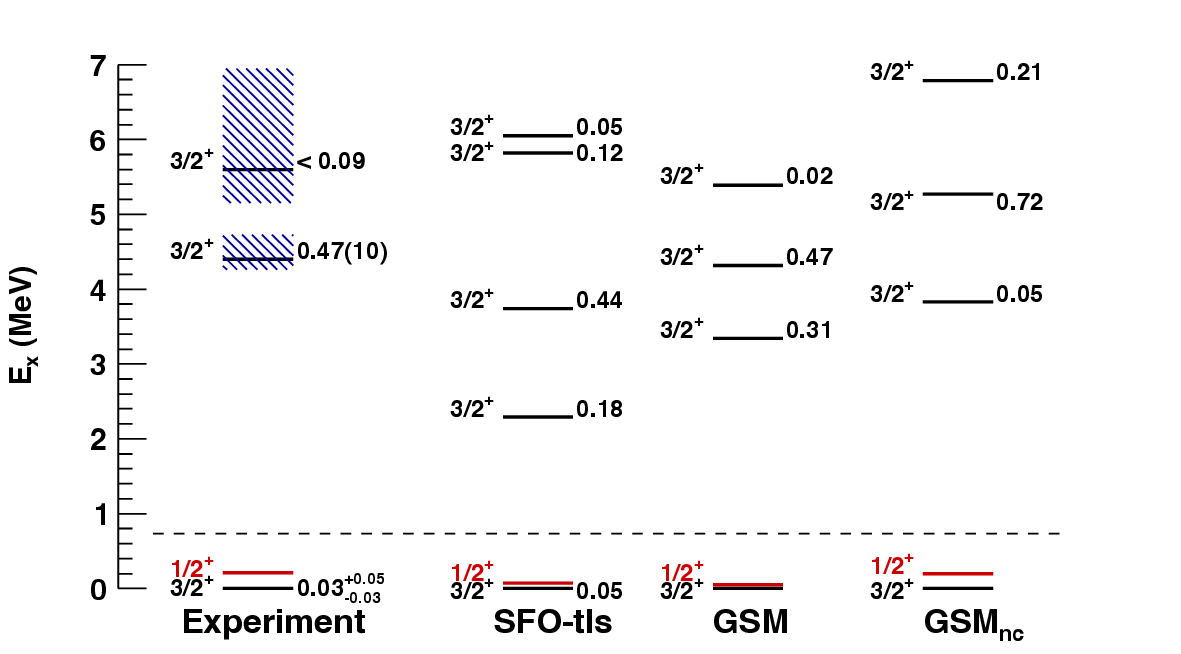}
\caption{\label{ex_nrj} Excitation energy ($E_x$) and spectroscopic factors ($C^2S(0^+_1)$) of the 3/2$^{+}$ states in $^{17}$C from the present work (unbound levels) and Ref.~\cite{XPereira} compared to shell-model calculations employing the SFO-tls interaction and Gamow Shell Model calculations with (GSM) and without (GSM$_{nc}$) coupling to the continuum. Hatched bands represent the experimental uncertainties on the energies (Table~\ref{tab_sf_17c}).  The dashed line indicates the $^{17}$C single-neutron separation energy. The location of the lowest 1/2$^{+}$ state with the largest  $C^2S(0^+_1)$ spectroscopic factor for the $1s_{1/2}$ orbital  is also indicated.}
\end{figure}

Figure~\ref{ex_nrj} displays the excitation energies and $C^2S(0^+)$ spectroscopic factors of the $^{17}$C 3/2$^+$ states compared to shell model calculations in the $p$-$sd$ configuration space using the SFO-tls interaction \cite{TSuzukiI,TSuzukiII} and to Gamow Shell Model (GSM) \cite{GSM1,GSM2} calculations.  The latter employed a $^{14}$C core together with the Furutani-Horiuchi-Tamagaki interaction \cite{Furutani1,Furutani2,Jaganathen} and included explicit continuum coupling and internucleon correlations in a unified description. In both cases, only 3/2$^{+}$ states below 7 MeV are plotted.  A detailed summary of the experimental results obtained here and those of both calculations are provided in Table~\ref{tab_sf_17c}.

Both approaches predict the major part of the 3/2$^+$ strength to be shared between two states in the region of 2 to 4 MeV, The shell model predicts the higher of the two levels to be over twice as strong as the lower level, whereas the GSM finds a much more equal sharing of the strength, with the levels more closely spaced and lying some 0.5 to 1~MeV higher in energy.  Experimentally the
strength is centred relatively close in energy to the strongest states in both calculations.  Whilst it may be tempting to identify the resonance
observed in the experiment at 4.4~MeV with the shell model level at 3.7~MeV and that of the GSM at 4.3~MeV, the very large observed width does not allow the possibility that it may be two (or more) overlapping resonances with some fraction of the strength localised in the region 
of 2 to 3~MeV to be discounted.  In the case of the higher lying resonance observed here, it may be noted that as opposed to the GSM, the shell model predicts a doublet with a strength similar to experiment.
It is interesting to note that the summed unbound 3/2$^+$ strength of both the shell model and GSM predictions is very similar (0.8).  
Experimentally it would appear, however, that a significant fraction of the 3/2$^+$ spectroscopic strength remains to be located and almost certainly lies at energies above 6~MeV which were not accessible in the present experiment.

In order to gain some insight into the effects of the continuum on the GSM predictions, the calculations have been repeated without the inclusion of the coupling to the continuum -- ``$nc$'' (Fig.~\ref{ex_nrj}).  Specifically, the GSM$_{nc}$ calculations were made using natural orbitals (pole-like orbitals) as compared to those with continuum coupling (GSM) which employed the standard Berggren basis in which bound, resonance and continuum single-particle states are treated on an equal footing in the complex momentum plane.  As may be be seen in Figure~\ref{ex_nrj}, the continuum coupling reduces the energies of the 3/2$^{+}$ resonances
and also shares the strength more evenly between the two lowest lying resonances.

\begin{figure}
\includegraphics[scale=0.4]{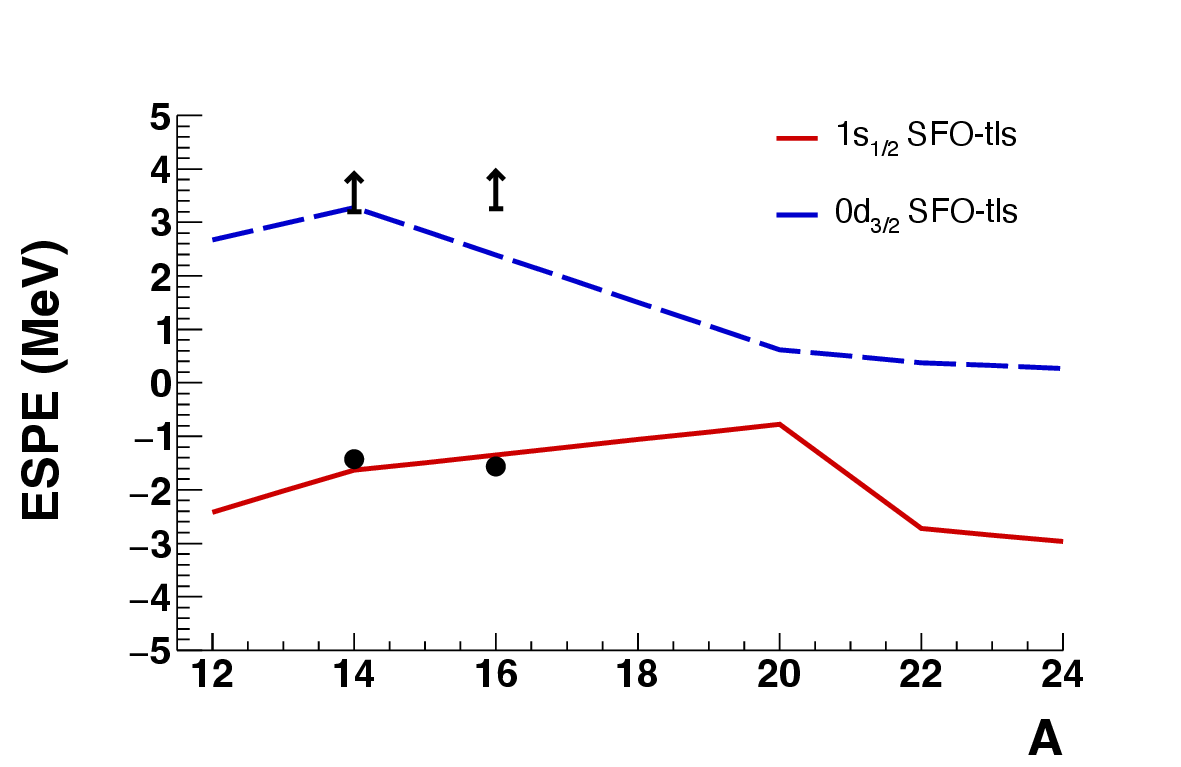}
\caption{\label{espe} Experimental effective single-particle energies (ESPE) for the $1s_{1/2}$ and $0d_{3/2}$  orbitals as derived from neutron addition on $^{16}$C in this work and Ref.~\cite{XPereira} and removal \cite{JLoisPLB} and from the literature (~\cite{ENSDF} for $^{14}$C. Comparison is made with the shell model for the SFO-tls interaction }
\end{figure}

In order to quantify the size of the N=16 shell gap, information from neutron stripping on $^{16}$C obtained here (unbound levels) and previously \cite{XPereira} (bound states) and from neutron pickup \cite{JLoisPLB} has been combined to deduce the effective single-particle energies (ESPE) \cite{Baranger} of the $\nu1s_{1/2}$ and $\nu0d_{3/2}$ orbitals.
Whilst for the $\nu1s_{1/2}$ orbital the strength is essentially exhausted by the $1/2^+_1$ state in $^{17}$C, the experimentally deduced spectroscopic factors for the $\nu0d_{3/2}$ orbital account for about half of the expected total strength ($\sum_{k=1}^{2} C^{2}S_{k}$= 0.50$^{+0.11}_{-0.10}$)\footnote{Only the C$^{2}$S(0$^{+}$) for the first two 3/2$^{+}$ states were considered: ground  \cite{XPereira} and first resonant state.}.  As such, the $\nu0d_{3/2}$ ESPE will thus represent a lower limit.
The ESPE obtained for the $\nu0d_{3/2}$ and $\nu1s_{1/2}$ orbitals are $\varepsilon_{\nu0d_{3/2}}$=3.40$^{+0.31}_{-0.14}$ MeV and $\varepsilon_{\nu1s_{1/2}}$=-1.68(30) MeV.
This leads to a value (lower limit) for the N=16 shell gap of  $\Delta$=$\varepsilon_{\nu0d_{3/2}}-\varepsilon_{\nu1s_{1/2}}$=5.08$^{+0.43}_{-0.33}$~MeV in $^{16}$C.  

Using the results available in the literature \cite{ENSDF} for the same reactions on $^{14}$C the ESPE could also be deduced.  That determined for the $\nu0d_{3/2}$ orbital represents a lower limit owing, as for $^{16}$C, to
the inability of experiment to locate all the associated spectroscopic strength.

\begin{table*}

\centering
\ra{1.4}
\caption{Excitation and resonances energies, $E_{x}$ and $E_{r}$ (MeV), corresponding transferred angular momenta $\ell$, total width $\Gamma^{tot}$ in MeV and spectroscopic factors $C^2S$ for the unbound states observed in $^{17}$C (assuming a two-level fit -- see text), compared to shell-model  predictions for the 3/2$^+$ levels with the SFO-tls interaction and to results from GSM calculations. Quantum numbers $n\ell j$ of the valence neutron configuration are indicated. The quoted uncertainties include statistical and fitting errors. }

\vspace{0.15cm}

\tabcolsep=0.11cm
\begin{tabular}{ccccccccccccccc}
\hline
\hline
\multicolumn{5}{c}{Exp}  &\multicolumn{6}{c}{SFO-tls}&\multicolumn{4}{c}{GSM}\\

\hline

$E_{x}$ & ${E_{r}}^{a)}$ &$\ell$ & $ \Gamma^{tot}$  & $C^{2}S(0^{+})$   &  $E_{x}$ & $E_{res}$ &$C^{2}S(0^{+})$  & \multicolumn{3}{c}{$C^{2}S(2_{1}^{+})$} &  $E_{x}$ & $E_{res}$ &$C^{2}S(0^{+})$  & \multicolumn{1}{c}{$C^{2}S(2_{1}^{+})$}\\

\cline{9-11}
 & & & &$0d_{3/2}$& & & $0d_{3/2}$  &$0d_{3/2}$&$0d_{5/2}$&$1s_{1/2}$  & & &$0d_{3/2}$ & $0d_{3/2}$\\
\hline
 -    & - & -  & &           & 2.29 & 1.55 & 0.18 & 0.09 & 0.04 & 0.15   &3.34 & 2.60 & 0.31 & 0.14\T \\
 -   &  - & -  & &           & 3.74 & 3.00 & 0.44 & 0.09 & 0.07 & 0.20 &  4.32 & 3.58 & 0.47 & 0.04\\
4.40$^{+0.33}_{-0.14}$ & 3.66$^{+0.33}_{-0.14}$ & 2  & 3.45$_{-0.78}^{+1.82}$  & 0.47(10) $^{b}$ &   & &  &  &  &  &  &  &  & \\
5.60$^{+1.35}_{-0.45}$ & 4.86$^{+1.35}_{-0.45}$  & 2  & 1.6$_{-1.4}^{+4.6}$ & $<$ 0.09 &          &  &  & &  & & & & &  \\
-   & -  & -  &  &         & 5.82 & 5.09 & 0.12 & 0.20 & 0.01 & 0.02  & 5.39 & 4.65 & 0.02 & 0.91\\
-   & -  & -  & - & -      & 6.05 & 5.32 & 0.05 & 0.34 & -      & -        & -      &  -       & - & -  \B\\
\hline
\hline
\label{tab_sf_17c}
\end{tabular}
\end{table*}%

\begin{table*}
\vspace{-0.7 cm}
\begin{tabular}{@{}p{15.9 cm}@{}}
 $^{a)}$ \footnotesize{For the decay to the $^{16}$C ground state.}\\
 $^{b)}$ \footnotesize{$C^{2}S$(0$^{+}$) = 0.45$_{-0.22}^{+0.32}$ from the branching ratio and decay width (see text).} \\
\end{tabular}
\end{table*}

As may be clearly seen in Figure~\ref{espe}, the N=16 gap in $^{16}$C remains comparable to that for $^{14}$C despite the decrease in the energy of the $0d_{3/2}$ orbital predicted by the SFO-tls interaction when adding two neutrons.  This is in contrast to the N=14 gap where configuration mixing results in its disappearance around $^{14-16}$C \cite{MStanoiu-N=14Cchain}. 
Compared to the values derived from the SFO-tls interaction, which are in line with the lower limit for the N=16 gap at $^{14}$C, it is at least 1.3~MeV larger in $^{16}$C than the prediction of 3.74 MeV.
The origin of this increase is beyond the scope of the present work but may be associated with the treatment of the continuum and/or the effects of deformation.  Such a deviation from the shell model predictions calls into question their validity as the dripline is approached and, as such, the size of the N=16 gap in $^{22}$C.

\section{Conclusions}

In summary, the d($^{16}$C,p) reaction has been employed to locate the $\nu 0d_{3/2}$ strength in$^{17}$C, which was found to lie at energies in excess of some 2  MeV above the single-neutron decay threshold.  Guided by theoretical expectations, the strength was determined to be primarily concentrated in a very broad resonance at around 4.4 MeV excitation energy. The proton backward angle differential cross sections, branching ratios for neutron decay to the $^{16}$C(2$_{1}^{+}$) level and partial decay widths permitted the spectroscopic strength to be deduced.  The measurements were found to be in reasonable agreement with shell model calculations using the SFO-tls interaction as well as Gamow Shell-Model calculations including continuum effects, although the theoretical strength was stronger in both calculations than that observed.   Using the experimentally deduced spectroscopic factors a lower limit could be placed on the ESPE of the neutron $0d_{3/2}$ orbital and thus the energy of the N=16 gap $-$ 5.08$^{+0.43}_{-0.33}~$ MeV$-$ which was at least 1.3~MeV larger than that derived from the shell model.
Given the predicted evolution of the gap as the carbon isotopes become more neutron-rich and the interest in $^{22}$C,
measurements of the $(d,p$) and $(d,t)$ reactions on $^{18}$C, although challenging, would be a welcome next step.

\section*{Acknowledgments}

The support provided by the technical staff of LPC-Caen and GANIL in preparing and executing the experiment is greatly appreciated.  J.L.F wishes to acknowledge financial support from Xunta de Galicia (Spain) grant number ED481A-2020/069, X.P.L. acknowledges support by IN2P3/CNRS (France) doctoral fellowship, the ST/P003885 grant (Spain) and Grant No. IBS-R031-D1. This work was supported by: the Spanish MINECO through the project PID2021-128487NB-I00, the Xunta de Galicia under project  2021-PG045 and the Maria de Maeztu Unit of Excellence MDM-2016-0692. W.N.C., N.T. and A.M. acknowledge financial support from the STFC grant number ST/\-L005743\-/1 and for ST/\-P005314/\-1. J.A.L, A.M.M and P.P acknowledge financial support from Grant No. PID2020-114687GB-I00 by MCIN/AEI/10.13039/501100011033. P.P. acknowledges  Ph.D. grant from the Ministerio de Universidades. M.F., N.C., Tz.K. and C.W acknowledge support from STFC grant number ST/\-V001043/\-1. The participants from the Universities of Birmingham and Surrey, as well as the INFN and IFIN-HH laboratories also acknowledge partial support from the European Community within the FP6 contract EURONS RII3-CT-2004-06065.

\appendix

\section{}
\label{SM_PS} 

\begin{figure}[h]
\includegraphics[scale=0.4]{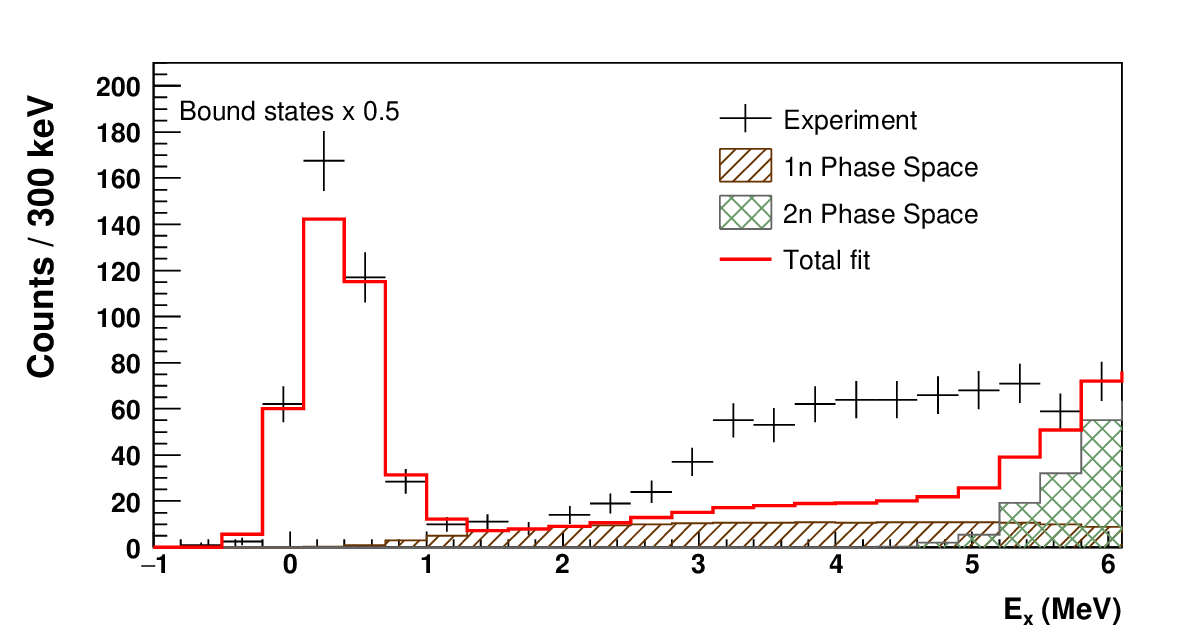}
\caption{\label{data_unbd_app} Reconstructed excitation energy of $^{17}$C for $\theta_{lab}$=[147$^{\circ}$, 167$^{\circ}$] where the maximised contributions from the one and two-neutron breakup channels (shaded histograms) are shown (see text). }
\end{figure}

The forms in excitation energy of the one and two-neutron breakup channels were estimated by sampling, with the restrictions imposed by the experimental acceptances, the three and four-body phase space distributions. Figure~\ref{data_unbd_app} shows the comparison with the experimental excitation energy spectrum whereby the contributions of each were maximised (in the regions, respectively, around 1.5 and 6 MeV) without exceeding at any energy the measurement. As such, it is clear that there is excess strength in the region of 2 to 5.5~MeV that must arise from continuum states of $^{17}$C.

\section{}
\label{SM_1R} 
\begin{figure}[h]
\includegraphics[scale=0.4]{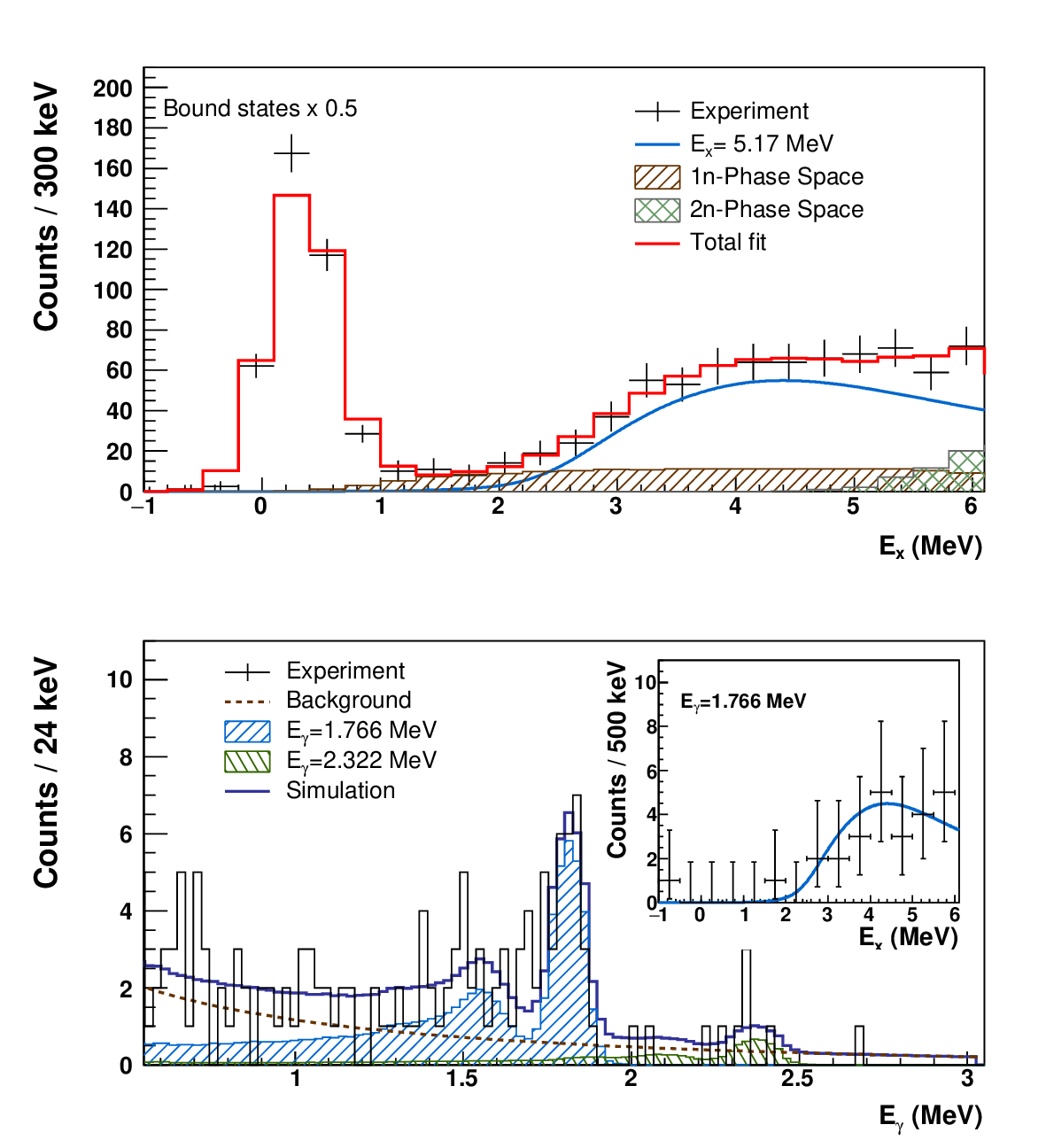}
\caption{\label{data_1r} Same panels as for Fig.~\ref{data_unbd} but with the inclusion of only a single resonance.}
\end{figure}

\begin{figure}[h]
\includegraphics[scale=0.35]{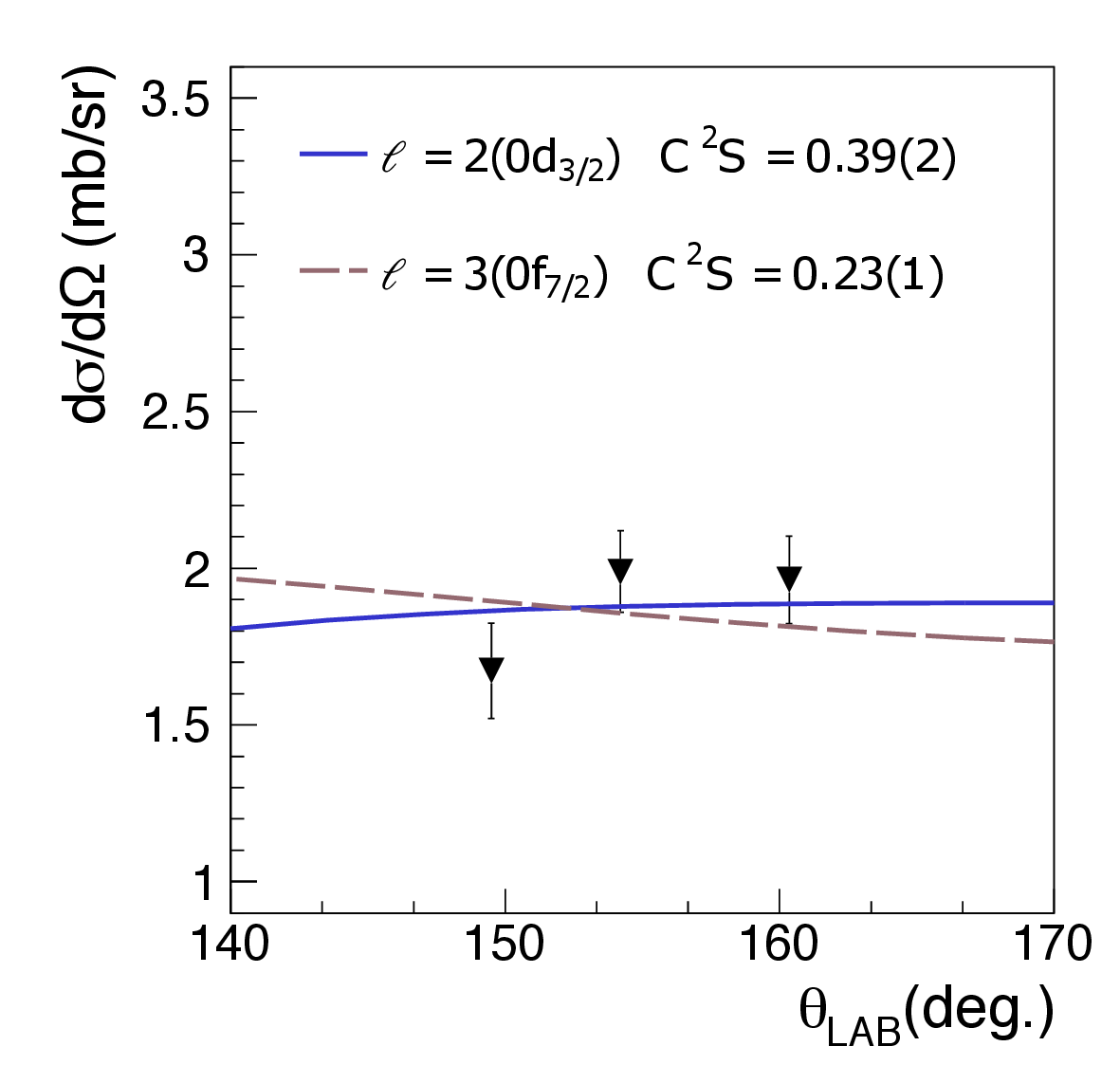}
\caption{\label{ang_dist_1r} Proton backward angle differential cross sections for the resonance (data points shown with statistical error bars) compared to ADWA calculations for transfer to the $\nu0d_{3/2}$ (solid blue line) and $\nu0f_{7/2}$ (dashed brown line) orbitals.} 
\end{figure}

The experimental excitation energy spectrum can be well described, in addition to the bound states and contributions from the 
one and two-neutron breakup channels, by the inclusion of only a single resonance.  The results of such a fit are shown in
Figure~\ref{data_1r} whereby the resonance is found at 
E$_{x}$=5.17$^{+1.53}_{-0.43}$~MeV with a total width of $\Gamma_{1}^{tot}$=6.2$^{+8.00}_{-1.9}$~MeV. The branching ratio for decay to the $^{16}$C(0$_{1}^{+}$) state was determined to be 0.28(14).  Given the single-particle width for $\ell$=2 decay at the resonance energy in question -- $\Gamma_{sp}=$4.24 MeV -- the corresponding spectroscopic factor was deduced to be $C^{2}S$(0$_{1}^{+}$)=0.41$^{+0.53}_{-0.12}$.  This may be compared to a value of 0.39(2) derived from the proton backward angle differential cross sections (Fig. \ref{ang_dist_1r})
As such, a lower limit for the ESPE of the $\nu0d_{3/2}$ orbital may be deduced 4.08$^{+1.62}_{-0.55}$.  The corresponding lower limit for the N=16 shell gap of 5.76$^{+1.63}_{-0.59}$~MeV is in accord, within uncertainties, with that determined based on a two-resonance fit.


\begin{thebibliography}{99}
  
\bibitem{MayerIII}
  {M. G. Mayer },
  {Phys. Rev.}
  {78}
  {(1950)}
  {16}.
  
\bibitem{MayerIV}
  {M. G. Mayer },  
  {Phys. Rev.}
  {78} 
  {(1950)}
  {22}.
  
 \bibitem{Hoffman-N16}
 {C. R. Hoffman et al.},
{Phys. Rev. Lett.}
{100} 
{(2008)}
 {152502}.
      
\bibitem{TshooI}
{K. Tshoo et al.},
{Phys. Rev. Lett.}
 {109} 
{(2012)}
{022501}.


 \bibitem{TOtsukaI}
 {T. Otsuka, R. Fujimoto, Y. Utsuno, B. Alex Brown, M. Honma and T. Mizusaki},
  {Phys. Rev. Lett.}
   {87} 
 {(2001)}
   {082502}.
 \bibitem{VMU} T. Otsuka et al., Phys. Rev. Lett. 104 (2010) 012501.

\bibitem{NSmirnova}
 {N. Smirnova, B. Bally, K. Heyde, F. Nowacki, K. Sieja},
  {Phys. Lett. B}
   {686} 
   {(2010)}
   {109}.

\bibitem{SorlinIII}
  {T. Otsuka, A. Gade, O. Sorlin, T. Suzuki and Y. Utsuno},
  {Rev. Mod. Phys.}
   {92} 
   {(2020)},
   {015002}.
 
   
\bibitem{Tanaka}
  {K. Tanaka et al.},
  {Phys. Rev. Lett.}
   {104} 
   {(2010)},
   {062701}.

\bibitem{Togano}
  {Y. Togano et al.},
  {Phys. Lett. B}
   {761} 
   {(2016)},
   {412}.
   
\bibitem{Nagahisa}
  {T. Nagahisa, W. Horiuchi},
  {Phys. Rev. C}
   {97} 
   {(2018)},
   {054614}.
   
\bibitem{Kobayashi}
  {N. Kobayashi et al.},  
  {Phys. Rev. C}
   {86} 
   {(2012)},
   {054604}.
   
\bibitem{Coraggio}
  {L. Coraggio et al.}, 
  {Phys. Rev. C}
   {81} 
   {(2010)},
   {064303}.
   
\bibitem{Sun}
   {X-X Sun, J. Zhao, S-G. Zhou},
  {Phys. Lett. C}
   {97} 
   {(2018)},
   {054614}.

\bibitem{Darden}  
   {S. E. Darden et al.},
   {Phys. Rev. C}
   {32} 
   {(1985)}
   {1764}.	
 
\bibitem{Stefan}
  {I. Stefan et al.},
  {Phys. Rev. C}
   {90} 
   {(2014)},
   {014307}.
  
\bibitem{XPereira}
  {X. Pereira-L\'opez et al.},
  {Phys. Lett. B}
   {811} 
   {(2020)}
   {135939}.

\bibitem{MStanoiu-N=14Cchain}
  {M. Stanoiu et al.},
  {Phys. Rev. C}
  {78} 
  {(2008)}
  {034315}.
  
\bibitem{Ueno} 
{H. Ueno et al.},
{Phys. Rev. C}
 {87}  
 {(2013)}
 {034316}.

\bibitem{Ogawa} 
{H. Ogawa et al.},
  {Phys. Lett. B}
 {451}  
 {(1999)}
 {11}.

\bibitem{Baumann} 
{T. Baummann et al.},
{Phys. Lett. B}
 {439}  
 {(1998)}
 {256}.

\bibitem{SauvanPLB} 
  {E. Sauvan et al.},
  {Phys. Lett. B}
   {491}
   {(2000)}
   {1}.
 
 \bibitem{SauvanPRC} 
  {E. Sauvan et al.},
  {Phys. Rev. C}
   {69}
   {(2004)}
   {044603}. 
   
\bibitem{Maddalena}
  {V. Maddalena et al.},
  {Phys. Rev. C}
   {63}
   {(2001)}
   {024613}.


\bibitem{UDatta}
  {U. Datta-Pramanik et al.},
  {Phys. Lett. B}
   {551} 
   {(2003)}
   {63}.

  
\bibitem{AME16}
  {M. Wang et al.},
  {Chinese Phys. C}
   {45} 
   {(2020)}
   {030003}.

\bibitem{Bohlen}
  {H.G. Bohlen et al.},
  {Eur. Phys. J. A},
   {31}
   {(2007)}
   {279}.
    


\bibitem{Satou}
  {Y. Satou et al.},
  {Phys. Lett. B}
   {660} 
   {(2008)}
   {320}.


\bibitem{Kim}
  {S. Kim et al.},
  {Phys. Lett. B},
   {836}
   {(2023)}
   {137629}.
   

\bibitem{LISE-bis}
  {R. Anne and A.C. Mueller et al.},
   {Nucl. Instrum. Meth. B}
   {70} 
   {(1992)}
   {276}.
  
\bibitem{TIARA}
  {M. Labiche et al.},
   {Nucl. Instrum. Meth. A}
   {614} 
   {(2010)}
   {439}.

\bibitem{EXOGAM}
  {J. Simpson et al.},
  {Acta Phys. Hung., New Series, Heavy Ion Physics},
   {11}  
   {(2000)}
   {159}.

\bibitem{CHARISSA}
  {N. I. Ashwood et al.},
  {Phys. Rev. C}
   {70}  
   {(2004)}
   {024608}.
   
\bibitem{XPereiraPhD}
  {X. Pereira-L\'opez, Study of transfer reactions induced by a $^{16}$C beam, PhD Thesis Universit\'e de Caen-Normandie and USC-Santiago (2016)},
  {http://hal.in2p3.fr/tel-01522695}
      
 \bibitem{JLoisPLB}
  {J. Lois-Fuentes et al.},
  {Phys. Lett. B}
   {845} 
   {(2023)}
   {138149}.

  \bibitem{BFD-O21}
  {B. Fern\'{a}ndez-Dom\'{i}nguez et al.},
  {Phys. Rev. C}
   {84}
   {(2011)}
   {011301}.

   
   \bibitem{GEANT4}
  {The GEANT4 Collaboration (S.~Agostinelli et al.)},
  {Nucl. Instrum. Meth. A}
   {506}  
   {(2003)}
   {250};
   {https://geant4.web.cern.ch/geant4/}   
   
   \bibitem{ENSDF}
  {ENSDF},  
 {https://www.nndc.bnl.gov/ensdf/}

\bibitem{LaneThomas}
  {A.M. Lane and R.G. Thomas},
  {Rev. Mod. Phys. }
   {30}  
   {(1958)}
   {257}.

\bibitem{dwuck}P. D. Kunz. http://spot.colorado.edu/$~$kunz/DWBA.html

\bibitem{KoningDelaroche}
  {A. J. Koning and J. P. Delaroche},
  {Nucl. Phys. A}
   {713}
   {(2003)}
   {231}.
   
\bibitem{vf} C. M. Vincent and H. T. Fortune, Phys. Rev. C {\bf 2}, (1970) 782 .
  
  \bibitem{TSuzukiI}
  {T. Suzuki, R. Fujimoto and T. Otsuka},
  {Phys. Rev. C}
   {67} 
   {(2003)}
   {044302}.

 \bibitem{TSuzukiII}
  {T. Suzuki and T. Otsuka},
  {Phys. Rev. C}
   {78} 
   {(2008)}
   {061301(R)}.
   
\bibitem{GSM1}  
   {N. Michel and M. P\l{}oszajczak},
   {Gamow Shell Model - The Unified Theory of Nuclear Structure and Reactions},
   {(Springer, Berlin)},
   {983} 
   {(2021)}.

     \bibitem{GSM2}  
   {N. Michel, W. Nazarewicz, M. P\l{}oszajczak and K. Bennaceur},
   {Phys. Rev. Lett.}
   {89} 
   {(2002)}
   {042502}.
  
    \bibitem{Furutani1}  
   {H. Furutani, H. Horiuchi and R. Tamagaki},
   {Prog. Theor. Exp. Phys.}
   {62} 
   {(1979)}
   {981}.
   
   \bibitem{Furutani2}  
   {H. Furutani et al.},
   {Prog. Theor. Phys. Supp}
   {68} 
   {(1980)}
   {193}.
   
\bibitem{Jaganathen}  
   {Y. Jaganathen et al.},
   {Phys. Rev. C}
   {96} 
   {(2017)}
   {054316}.	
   
\bibitem{Baranger}
	 {M. Baranger,} 
	 {Nucl. Phys. A}
	 {149} 
	 {(1970)} 
	 {225}.
	 
	

\end{thebibliography}
\end{document}